\documentstyle[11pt,psfig]{article}

\newcommand\etal{{\it et al.}}
\begin{document}
\begin{titlepage}
\begin{center}

{ \large \bf Prototype Detector for Ultrahigh Energy Neutrino
Detection$^{\dagger}$}
\renewcommand{\thefootnote}{\fnsymbol{footnote}}\footnotetext{$^{\dagger}$
Submitted to {\it Astroparticle Physics}}

\vspace{.3in}

{\bf Joshua R. Klein$^*$
\renewcommand{\thefootnote}{\fnsymbol{footnote}}\footnotetext{$^*$
jrk@upenn5.hep.upenn.edu} and Alfred K. Mann$^{**}$
\renewcommand{\thefootnote}{\fnsymbol{footnote}}\footnotetext{$^{**}$
mann@dept.physics.upenn.edu}\\
Department of Physics \& Astronomy\\
University of Pennsylvania$^{\ddagger}$
\renewcommand{\thefootnote}{\fnsymbol{footnote}}\footnotetext{$^{\ddagger}$
Supported in part by the U.S. Department of Energy}\\
Philadelphia, PA 19104

\vspace{.3in}

May 27, 1998}

\vspace{1.3in}

Abstract

\end{center}

\vspace{.3in}

Necessary technical experience is being gained from successful construction and
deployment of current prototype detectors to search for UHE neutrinos in
Antarctica, Lake Baikal in Russia, and the Mediterranean. 
The
prototype detectors have also the important central purpose of determining whether or not UHE neutrinos do
in fact exist in nature by observation of at least a few UHE
neutrino-induced leptons with properties that are not consistent with expected
backgrounds. We discuss here the criteria for a prototype detector to accomplish
that purpose in a convincing way even if the UHE neutrino flux is
substantially lower than predicted at present.
\end{titlepage}

\noindent {\bf Introduction}

\vspace{1ex}

There are several current efforts to construct deeply buried particle
detectors of very large dimensions with which to search for ultrahigh energy (UHE)
neutrinos from space [1]. These efforts involve prototype detectors aimed at
mastering equipment design and techniques for
deployment in ice or in deep water. The sensing elements to detect the leptonic
products of  UHE neutrino interactions in the vicinity of the detector are
photomultiplier tubes (PMTs) and their associated circuitry which present new
technical problems of power capacity and distribution and of data
acquisition
 insofar as the PMTs are deployed at a large distance from the power source
and
 in an unusual medium.
Deployment in ice or in the sea, particularly the deep sea, requires
development of new designs[2],[3] and of an infrastructure new to particle physicists. These issues need
to be studied empirically by the experience that the prototype detectors are
meant to provide.

However, the prototype detectors have an important physics purpose in addition to
answering the technical questions above. This purpose is to determine whether or
not
the hypothesized  UHE neutrino sources which are the object of
the search do in fact exist in nature. Specifically, the purpose of the
prototype detectors  is to demonstrate the existence of UHE neutrinos by
observation of at least a small number of UHE neutrino-induced muons or
neutrino-induced electrons with properties that are not consistent with
expected, well-understood backgrounds. It is difficult to foresee the
construction of a detector much larger than a prototype detector in the
absence of such a
 proof of existence.

In this note we discuss briefly the criteria to be satisfied by a prototype
detector to accomplish that purpose in a convincing way. We rely heavily on
the valuable, encyclopedic paper of Gandhi et al.[4], but concentrate
specifically on the criteria necessary to achieve an UHE neutrino-induced
signal above background subject to the perhaps pessimistic assumption that the
sought-for UHE neutrino flux is an order of magnitude lower in intensity than
the current predicted values. This is not a mindless assumption because the
UHE neutrino flux calculations are strongly dependent on the uncorroborated
models chosen to simulate the acceleration mechanisms in extragalactic and
cosmic sources. A lower than predicted UHE neutrino flux would be similar in
intensity to known backgrounds and difficult to extract convincingly from them
without rethinking how the search should be performed and how a useful upper
limit on the neutrino flux can be obtained. Our aim is to suggest a minimal
detector and to indicate how the location and operation of the detector will
discriminate against backgrounds and provide a high probability of observing a
few UHE neutrino-induced leptons in one or two years of exposure even if
the actual UHE neutrino flux is as much as a factor of ten lower than the
predicted values.
\vspace{2ex}

\noindent{\bf UHE Neutrino-Induced Muon Rates and Angular Distributions}

\vspace{1ex}

Interest in a search for UHE neutrinos from space lies in the possibility that
astronomical or cosmological sources might give rise to neutrinos in the
energy region between $10^6$ GeV and $10^{12}$ GeV. That is the
energy region in which attempts to model power generating mechanisms involving
black holes and accretion disks in active galactic nuclei (AGN) yield
observable UHE neutrinos fluxes [5], as do also models of relativistic
fireballs for gamma-ray bursters (GRBs) [6]. Various conjectures in elementary
particle and cosmological theory introduce so-called topological defects (TDs)
that give rise to the decays of very massive remnant particles, one of whose
decay products might be an UHE neutrino [7]. A summary of the results of these
model-based calculations of neutrino fluxes---which we take as order of
magnitude estimates---is given in Fig. 1.

It is realistic to consider the prospect of observing the low intensity fluxes
in Fig. 1 because the reaction cross section for neutrino plus nucleon rises
steeply with neutrino energy as shown in Fig. 2 for both neutrinos and
antineutrinos [4]. Observe that the cross section rises by roughly six decades as
the neutrino energy increases by eight decades from 1 GeV to $10^8$ GeV,
despite the production of real intermediate vector bosons as indicated by the
breaks in the curves in Fig. 2 at about $10^5$ GeV.

A further consequence of the rapid increase of neutrino interaction cross
sections is that the Earth becomes a significant absorber of neutrinos above a
certain neutrino energy, roughly $10^4$ GeV, and is essentially opaque to
neutrinos with energy above about $10^7$ GeV. The neutrino survival
probability as a function of $\cos \theta_Z$, $\theta_Z$ the zenith angle at
the detector, for three neutrino energies is plotted in Fig. 3. This 
 well-understood effect is
particularly damaging because the  angular region in which one naturally
thinks to search for leptons from UHE neutrino interactions near or in a
massive detector is  the  region $\cos \theta_Z \stackrel {<}{\sim} 0$.
 This effect requires some
 revision of most current plans for UHE neutrino searches, particularly if they
are to search successfully for UHE neutrino interaction products from a
neutrino flux significantly lower than is shown in Fig. 1.

	To estimate the differential flux $d \phi_{\mu}^{\nu}/d\Omega$ of 
UHE neutrino-produced muons from neutrinos in the energy interval $10^7$ to
$10^{10}$ GeV reaching a detector at a depth of 5 km.w.e., we 
numerically integrate over  neutrino energy and all possible 
neutrino interaction points in the Earth ($X$), weighted by the neutrino 
interaction probability and the probability of muon survival over a distance
from $X$ to the detector position ($D(\cos \theta_z)$):

\begin{equation}
\frac{d\phi_{\mu}^{\nu}}{d\Omega} = \int \frac{d\phi_\nu}{dE_\nu 
d\Omega}\int_0^{D(\cos\theta_z)} \sigma_{\nu N}(E_\nu) n
 e^{-\sigma_{\nu N}(E_\nu)nX} \int P^\mu_{\rm surv}(E_\mu,D(\cos\theta_z)-X)
(\frac{1}{\sigma_{\nu N}}\frac{d\sigma_{\nu N}}{d E_{\mu}}) dE_{\mu} dX dE_{\nu}
\label{fullint}
\end{equation}
We choose the representative UHE neutrino energy interval $10^7$ to $10^{10}$
because $10^7$ GeV is roughly the lowest energy at which it is likely that new
physics can be convincingly identified by the experimental method discussed
here and, on the other hand, observation of muons from $10^{10}$ GeV neutrinos
would unquestionably signify new physics.
We use 
the values calculated by Gandhi~\etal [4] for the total neutrino-nucleon
cross section $\sigma_{\nu N} (E_{\nu})$   and the Monte Carlo results of Lipari and 
Stanev [8] for the muon survival 
probability $P^{\mu}_{\rm surv}$.  The differential neutrino flux is taken from the values
shown in Fig.~1.  For simplicity, we replace the differential
cross section for muon production with a delta function centered on half
the incident neutrino energy, i.e.,
\[
\frac{1}{\sigma_{\nu N}}\frac{d\sigma_{\nu N}}{d E_{\mu}} =
 \delta(E_{\mu} - \frac{1}{2}E_{\nu}) 
\]
which reduces Equation~\ref{fullint} to
\begin{equation}
\frac{d\phi_{\mu}^{\nu}}{d\Omega} = \int \frac{d\phi_\nu}{dE_\nu d\Omega}\int_0^{D(\cos\theta_z)} \sigma_{\nu N}(E_\nu) n e^{-\sigma_{\nu N}(E_\nu)nX} P^\mu_{\rm surv}(\frac{E_\nu}{2},D(\cos\theta_z)-X) dX dE_{\nu}.
\label{redint}
\end{equation}
This substitution is adequate for the
order of magnitude estimates we present here.  Implicit 
in Equation~\ref{fullint} is the assumption that the muons are produced 
collinearly with the neutrinos, that is, $d \sigma_{\nu N}/d \Omega$ is a 
delta function centered on the incident neutrino direction.

In Figure 4a, we plot the UHE neutrino-induced muon flux as
a function of detector zenith angle $\cos\theta_z$, for a detector 5 km deep
and  for three different
incident neutrino energies to give an idea of the energy dependence within the
interval $10^7$ to $10^{10}$ GeV. As is evident, 
there are three distinct regions for each curve.  For 
$\cos\theta_z < -0.10$, the muon flux drops off rapidly---a consequence of the 
extreme attenuation of the neutrino flux shown in 
Fig. 3.  For $\cos\theta_z > 0.25$, the flux drops somewhat less 
rapidly because the neutrino target thins, resulting in fewer 
neutrino interactions and fewer muons.  Between those  limits, 
 the neutrino target thickness is small compared to the 
neutrino interaction length but greater than the muon range.  Thus the UHE
neutrino flux is unattenuated, and all neutrino interactions closer to the detector 
than the muon range yield detected muons.  This maximum detected muon flux 
does not change until the neutrino target thickness becomes smaller than the muon range,
at  $\cos\theta_z > 0.25$.

	These general features, which are largely the result of the geometry of the
experimental arrangement, can be confirmed with an even simpler model.
If we take the muon range at these energies to be roughly 20 km.w.e., then the 
only neutrino interactions that produce detectable muons occur within 
20 km.w.e. of the detector.  The total number of neutrino interactions in a 
target of thickness $D$ is just $\phi_{\nu} {\rm [1 - exp} (- \sigma_{\nu} N
(D))]$, of 
which only $20 {\rm km.w.e.}/D$ produce detectable muons.  So, for example, for a
neutrino flux at $10^7$~GeV of $10^{-4} {\rm (km^{-2} s^{-1} sr^{-1)}}$, and a cross 
section of $10^{-33} {\rm cm^2}$, we find at $\cos\theta_z = 0.1$ a muon flux
of $\sim 3 \times 10^{-7} {\rm (km^{-2} s^{-1} sr^{-1})}$, in good agreement with
the full integral calculation above.   In this model, for $\cos\theta_z > 0.25$
the flux drops as $ {\rm [1 - exp} (- \sigma_{\nu} N (D))]/ 
{\rm [1 - exp} (- \sigma_{\nu} N (20))] $
as the maximum distance to the detector becomes less than the 20 km.w.e 
range. 

	In either of these models, the width of the region of maximum
detected muon flux is a strong function of detector depth. We plot
 in Figure 4b the angular distribution of the UHE neutrino-induced  
 muon flux as a function of $\cos\theta_z$ for a detector at a depth 
of 2 km.w.e., which shows a steeper fall-off of the acceptance region at small
zenith angles as the detector moves closer to the surface. To obtain the
relative UHE neutrino-induced muon intensities for the two detector depths
in Fig. 4, it is necessary to take into account the angular limits imposed by
the large cosmic ray muon flux, which are different for the different detector
depths as discussed in the next section. These limits are indicated in Fig. 5
with the UHE neutrino-induced fluxes integrated over the energy interval
$10^7$ to $10^{10}$ GeV for each of the detector depths. Summing the
UHE-induced muon flux from the cosmic ray muon limits to $\cos\theta_z \simeq
-0.10$ yields the UHE neutrino-induced muon flux at a detector depth of 5 km
integrated over the solid angle acceptance of the detector to be approximately
$ I^{\nu}_{\mu} = 5.3 \times 10^{-6} {\rm km^{-2} sec^{-1}}$, and the ratio of
the induced muon fluxes at 5 km and 2 km to be 1.7.
	
\vspace{2ex}

\noindent{\bf Backgrounds}

\vspace{1ex}

Apart from the serious technical difficulties involved in working deep under
ice or under the sea, the primary obstacle to achieving a convincing
demonstration of the observation of an UHE muon
from an UHE neutrino source in space is background from two known sources of
energetic muons; cosmic ray muons in Fig. 5 coming directly from primary cosmic ray
proton (and heavier element) interactions in the Earth's atmosphere, and muons
produced by the cosmic ray (often called atmospheric) neutrinos that are the
birth companions of the cosmic ray muons. These empirically well-studied
backgrounds fall off rapidly with increasing muon energy, but their energy
distributions have long tails extending past $10^7$ GeV. This is shown for the
cosmic ray muons in Fig. 6 [9] which is a companion to the muon plots in Fig.~5, and for the
atmospheric (atmos) neutrinos (not muons) in Fig. 1 [10].

\vspace{2ex}

\noindent{\bf Cosmic Ray Muon Background}

\vspace{1ex}

As indicated in Fig. 4, the zenith angular interval in which to search
efficiently for an UHE neutrino signal in a 5 km deep detector is $-0.10
\stackrel{<}{\sim} \cos \theta_Z \stackrel{<}{\sim} 0.25$.
Cosmic ray muons reaching the detector with $\cos \theta_Z \stackrel{<}{\sim}
0.25$ must first traverse at least 20 km w.e.; this requires an incident
cosmic ray muon energy of about $10^8$ GeV for a muon detector energy
threshold of $10^3$ GeV, and $10^9$ GeV for a muon detector threshold of $10^4$
GeV. For either energy threshold the cosmic ray muon flux satisfying the angle
and energy criteria is roughly $10^{-17} {\rm (km^{-2} sec^{-1} sr^{-1})}$.
For comparison, note that cosmic ray muons reaching a 2 km deep detector with
$\cos \theta_Z \stackrel{<}{\sim}  0.25$ need only traverse 8 km w.e., which
requires an incident muon energy of $10^6$ GeV for a detector energy threshold
of $10^4$ GeV, and is satisfied by the much  larger cosmic ray muon flux of about $ 5 \times
10^{-7} {\rm (km^{-2} sec^{-1} sr^{-1})}$.  As a consequence, the angular
limit to the acceptance region of a 2 km deep detector is approximately at
$\cos \theta_Z = +0.10$, where the cosmic ray muon flux is equal to that for a
5 km deep detector at the angular limit $\cos \theta_Z = +0.25$. These are the
cosmic ray muon limits shown in Fig. 5.

Although the cosmic ray muon flux is small beyond the angular limits in Fig.
5, the steep rise of the flux places a significant demand on the angular
resolution of the detector at either depth. To emphasize this fact, the cosmic
ray muon fluxes plotted in Fig. 5 are integrated over all energies equal to
and above $10^5$ GeV. One can see graphically that a large, possibly
overestimated, lower energy muon background is latent at angles just slightly
below the limiting angles shown. For example, a 5 km deep
detector with a muon energy threshold of $10^4$ GeV accepting muons at an
angle given by $ \cos \theta_Z = 0.30$ will measure a flux of $10^8$ GeV
 cosmic ray muons
 roughly 100 times larger
than the flux at $\cos \theta_Z = 0.25$, but still small relative to the UHE
neutrino-induced muon flux. However, at $\cos \theta_Z = 0.30$, there are
approximately 10 times as many lower energy cosmic ray muons as UHE neutrino-induced
muons so that the low energy tail on the cosmic ray muon energy  distribution
will require the 10 TeV detector threshold to eliminate them.
On the other hand, angular cuts that are more conservative than those in Fig. 5
clearly eat away at the useful signal. Analogous statements can be made
perhaps with greater force for a 2 km deep detector.

\vspace{2ex}

\noindent{\bf Atmospheric Neutrino-Induced Muon Background}

\vspace{1ex}

The atmospheric neutrino-induced muon background is more difficult to subdue.
In the interval $10^5$ to $10^6$ GeV---a representative energy interval for
atmospheric neutrinos (see below)---they produce a muon yield one half as
large as the muon yield from UHE
neutrinos in the energy interval $10^7$ to $10^{10}$ GeV. 

However, the different muon energy distributions from the two sources allow
them to be differentiated by the 10 TeV energy threshold of the detector.
Along the path to the detector, all of the muons from $10^5$ GeV neutrino
interactions in the 2 km segment closest to the detector will pass the energy
threshold, as will approximately 2/3 of the muons from the segment of path 2
to 4 km from the detector, and 1/3 from the segment 4 to 6 km away. Few
atmospheric neutrino-induced muons originating beyond 6 km will be detected.
This is illustrated in Fig. 7, taken from reference [8], which shows the
energy distributions of $10^5$ GeV muons after traversing different lengths of
path.

The average energy loss of muons is generally written as $ - dE_{\mu}/dx =
\alpha + \beta E_{\mu}$, with $\beta$ the sum of the fractional radiation
losses, which in the high energy limit are proportional to muon energy. In the
muon energy region between about $10^4$ to $10^9$ GeV,  $\beta$ is weakly
dependent on muon energy as indicated in Fig. 8, also taken directly from
reference [8]. The result is that the energy loss per unit path length of UHE
neutrino-induced $10^8$ GeV muons is closely $10^3$ times larger than that for the
atmospheric-induced $10^5$ GeV muons, which in turn means that the shape of
the energy distributions in Fig. 7 can be directly scaled to $10^8$ GeV
initial muon energy. Consequently, 95 percent of the muons produced by the UHE
$10^8$ GeV neutrinos in the distant segment of path between 8 and 10 km will
clear the threshold and a substantial fraction of the muons from 11 to 20 km
would also do so.
  
 The net result of the 10 TeV muon energy threshold is to reduce the detected
atmospheric neutrino-induced muon flux by a factor of approximately 5 and
yield an UHE signal about 10 times larger than the atmospheric background.

\vspace{2ex}

\noindent{\bf Muon Angle and Energy Resolution}

\vspace{1ex}

For the muon selection criteria to be effective and discriminate successfully
against the cosmic ray muon background, the angular resolution, $\Delta \cos
\theta_Z/\cos \theta_Z$, of a 5 km deep detector needs to be no worse than 0.2 in
the vicinity of $\cos \theta_Z = 0.25$ and, correspondingly, better than 0.5
for a 2 km deep detector at $\cos \theta_z \cong 0.10$. These requirements
are significant factors in the design of a prototype detector. For example,
assuming a cubic array of photomultiplier tubes (PMTs), the length of an edge
of the array and the PMT spacing within the array need to be optimized to
achieve the angular resolutions above, while attempting to obtain the area of
a side about ${\rm 0.1 km^2}$ and keeping the cost and deployment problems
realistic. This angular resolution is also satisfactory for the roughly
isotropic
angular distribution of the atmospheric neutrino background.

Discrimination against atmospheric neutrino-induced muons relies, as we have
seen, on the muon energy threshold. The muon yield from $10^4$ GeV atmospheric
neutrinos is 70 times larger than the muon yield from $10^5$ GeV atmospheric
neutrinos. However, the larger muon flux is, on average, below the 10 TeV
detector energy threshold even at production and below 1 TeV at the detector
entrance. Consequently, the rapidly rising atmospheric neutrino intensity with
decreasing neutrino energy is neutralized by the 10 TeV muon threshold.
Furthermore, muons from $10^6$ GeV atmospheric neutrinos are 200 times fewer
than the muons from $10^5$ GeV atmospheric neutrinos.

Accordingly, if the UHE neutrino flux is as large as indicated by present
estimates, neither the cosmic ray muon background nor the atmospheric
neutrino-induced muon background places an important requirement on the
detector energy resolution beyond the  10 TeV threshold cut. If, however,
the UHE neutrino flux is as much as an order of magnitude less than
anticipated,  demands on the energy resolution would become significant.
Discrimination against cosmic ray muons by measurement of their incident angle
would continue to be sufficient, but a tighter energy criterion might
lessen
the burden on the angular resolution. On the other hand, a lower UHE neutrino
flux would lead to a numerical ratio of UHE neutrino signal to atmospheric
neutrino background about equal to unity and require additional
discrimination that might be provided by selection criteria reflecting the
different energy distributions of the two muon samples as measured within the
detector.  Detailed study of the
observed event distributions, reinforced by Monte Carlo simulations, are
likely to
allow for statistical separation of the two event samples, but require more
detail than is appropriate for this general discussion.

\vspace{2ex}

\noindent {\bf Summary and Conclusions}

\vspace{1ex}

To bring to fruition the idea of constructing a detector of sufficient size
and capability for an intensive search and detailed study of UHE neutrinos
from any source  requires first that the existence of such neutrinos be demonstrated
conclusively in a smaller, but still quite ambitious detector. The prototype detectors
now under design or under construction are directed at that goal, as well as at
the task of learning to deploy apparatus and operate  in the hostile
environments of deep ice or the deep sea.

In this paper we have suggested three correlated operational requirements that
a
prototype  detector should satisfy to achieve
 that primary physics purpose even if the intensity of the sought-for UHE
neutrinos is an order of magnitude less than estimated at present.  First, the prototype
detector should be located under as much vertical shielding as possible,
preferably about 5 km. Second, analysis of the observed muons should
concentrate in the angular region $-0.10 \stackrel{<}{\sim} \cos \theta_Z
\stackrel{<}{\sim} 0.25$, $\phi = 2 \pi$, where the  UHE neutrinos
are not strongly absorbed by the Earth before they produce muons able to reach
the detector, and the produced UHE muons will not be
 outnumbered by the cosmic ray
muon flux. Indeed, the vertical depth of the detector provides a slant depth
of 20 km at $\cos \theta_Z = 0.25$ which ensures that the cosmic ray muon
background in the interval $-0.10 \stackrel{<}{\sim} \cos \theta_Z
\stackrel{<}{\sim} 0.25$ is negligible compared with the pessimistic estimate
of the magnitude of the UHE signal we have adopted here, providing the
detector angular resolution is adequate.
Third, in view of the high energy of the region, $10^6$ to $10^{12}$ GeV, in which
the search is best conducted and the rapid falloff with energy of the atmospheric
neutrino flux, the imposition of a 10 or possibly 20 TeV muon threshold in the data
analysis will maintain a numerical ratio of the reduced level of anticipated
signal to  atmospheric neutrino-induced background to 
of order unity. To improve this level of discrimination will require
statistical separation of the two components of the observing UHE muon energy
distribution, which might be achieved in a cubic detector array with linear
dimension approximately 0.3 km to absorb a statistically useful sample of
energy deposited by the transiting muon. Similar requirements apply to
neutrino-induced electrons.

We have given only rough
estimated absolute UHE muon signal rates based on the neutrino fluxes
in Fig. 1 or fluxes reduced by an order of magnitude because of the large uncertainties
involved. Our estimates of the UHE muon signal and the combined 
background from cosmic ray muons and atmospheric-induced muons suggest, however,
that 
observation of 3 to 10 UHE muons per year with energy greater than $10^6$ GeV
in a detector with geometric area of 0.1 km$^2$ is not unlikely, and that
these muons would not be confused with background. Such an observation would
clearly demonstrate the existence of one or more sources of UHE neutrinos in
space or of cosmological  origin and justify the construction of a ten or more
times larger area detector for detailed study of the new phenomenon.

\vspace{2ex}

\noindent {\bf Acknowledgements}

\vspace{1ex}

We are grateful to Todor Stanev for a particularly useful muon range table and a
number of discussions, and to T.K. Gaisser and David R. Nygren for
discussions. David N. Schramm, always interested in possible new phenomena,
was a source of encouragement.
 Some of these ideas crystallized for one of us (AKM) as a result
of witnessing the deployment exercise of a part of the Nestor detector in the
Ionian Sea in May, 1997.

\vspace{2ex}

\noindent {\bf References}

\vspace{1ex}

\begin{enumerate}

\item[] 1. Antarctica: the Antarctic Muon and Neutron Detector (AMANDA
Collaboration) is described by R. Morse, in Neutrino Telescopes, Proceedings
of the Fifth International Workshop, Venice, March 1993, edited by
M.~Baldo~Ceolin, University of Padua (1993), and by R. J. Wilkes in
Proceedings of the 22nd Annual SLAC Summer Institute on Particle Physics,
Stanford, CA (1994); P. Askebjer {\it et al.}, Science {\bf 267} (1995) 1147.

 In the Mediterranean: Proposal to the IN2P3 Scientific Committee by
ANTARES Collaboration (1996).

Also in the Mediterranean: the NESTOR project is
described by L.~K.~Resvanis in Neutrino Telescopes, Proceedings of the Fifth
International Workshop, Venice, March 1993, edited by M.~Baldo~Ceolin,
University of Padua (1993), and in the Proceedings of the Third NESTOR
International Workshop, October, 1993, edited by L.~K.~Resvanis, Athens
University Press (1994).

Lake Baikal, Russia: The Baikal Neutrino Telescope is described by
 R.~Wischnewski in the Proceedings of the Third NESTOR International Workshop,
October, 1993, edited by L.~K.~Resvanis, Athens University Press (1994); I.
Sokalski (for the Baikal Experiment) in Proc. of the Int'l Workshop XXXIInd
Rencontres de Moriond, January, 1997.

\item[] 2. The earliest effort to construct a large area neutrino detector in
the deep sea from which much was learned was the Deep Underwater Muon and
Neutrino Detector (DUMAND). This detector is no longer in operation. It is
described in J.~G.~Learned, Phil. Trans. Roy. Soc. London A 346 (1994).

\item[] 3. For novel detector developments, see D.R. Nygren {\it et al.}
LBL-38321, UC 412.

\item[] 4. R. Gandhi, C. Quigg, M.H. Reno, and I.~Sarcevic, Astropart. Phys.
{\bf 5} (1996) 81; G.C. Hill, Astropart. Phys. {\bf 6}  (1997) 215.

\item[] 5. Recent estimates of neutrino fluxes from AGN are: L. Nellen, K.
Mannheim, and P.~L.~Biermann, Phys. Rev. {\bf D47} (1993) 5270; A.P. Szabo and
R.J. Protheroe, Astropart. Phys. {\bf 2} (1994) 375; K. Mannheim, Astropart.
Phys. {\bf 3} (1995) 295; D. Kazanas in Proceedings of the Third NESTOR
International Workshop, October 1993, edited by L.K. Resvanis (Athens
University Press) 1994; F.W. Stecker and M.H. Salamon, Space Science Reviews
{\bf 75} (1996) 341.

\item[] 6. E. Waxman and J. Bahcall, Phys. Rev. Lett. {\bf 78} (1997) 2292.

\item[] 7.  C.T. Hill, D.N. Schramm and T.P. Walker, Phys. Rev. {\bf D36} (1987)
1007; J.H. MacGibbon and R.H. Brandenberger, Nucl. Phys. {\bf B331} (1990)
153; P. Battacharjee, C.T. Hill and D.N. Schramm, Phys. Rev. Lett. {\bf 69}
(1992) 567; S. Yoshida and M. Teshima, Prog. Theor. Phys. (Kyoto) {\bf 89}
(1993) 833; G. Sigl, S. Lee, D.N. Schramm and P. Coppi, Phys. Lett. B (1996);
R.J. Protheroe and P.A. Johnson, Astropart. Phys {\bf 4} (1996) 253.

\item[] 8. P. Lipari and T. Stanev, Phys. Rev. {\bf D44} (1991) 3543.

\item[] 9. S. Hayakawa, Cosmic Ray Physics, John Wiley and Sons, New York
(1969).

\item[] 10. V. Agrawal, T.K. Gaisser, P. Lipari and T.Stanev, Phys. Rev. {\bf D53}
(1996) 1314; T.K. Gaisser and T. Stanev, Proc. 24th Int'l Cosmic Ray
Conference (Rome, 1995) Vol. 1, p. 694.

\end{enumerate}

\newpage

\noindent{\bf Figures}

\begin{enumerate}

\psfig{figure=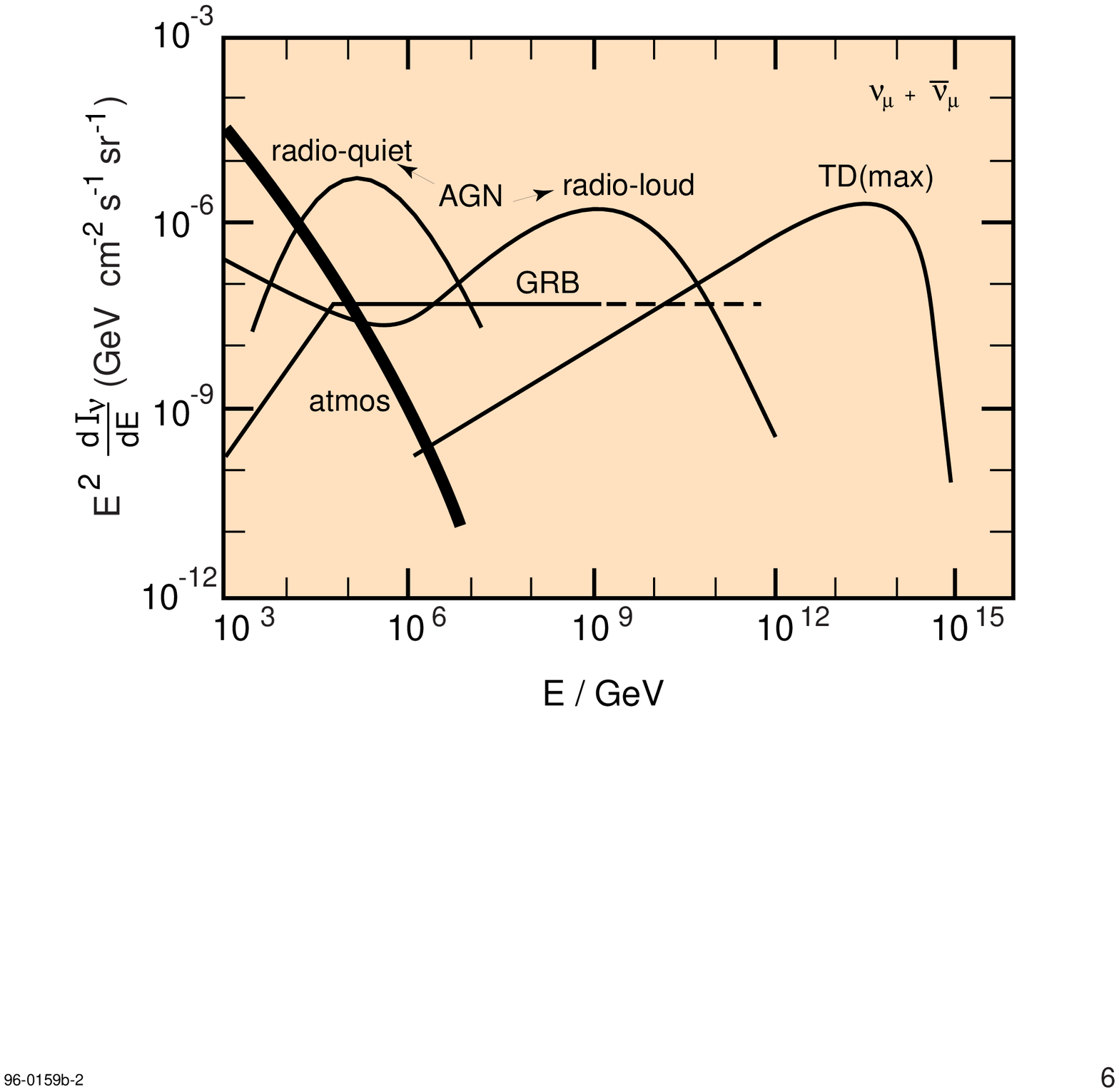,height=7in}
\item[] Fig. 1. Calculated UHE neutrino plus antineutrino fluxes at the
Earth's surface weighted by $E^2$. See text and reference 5.
\newpage

\psfig{figure=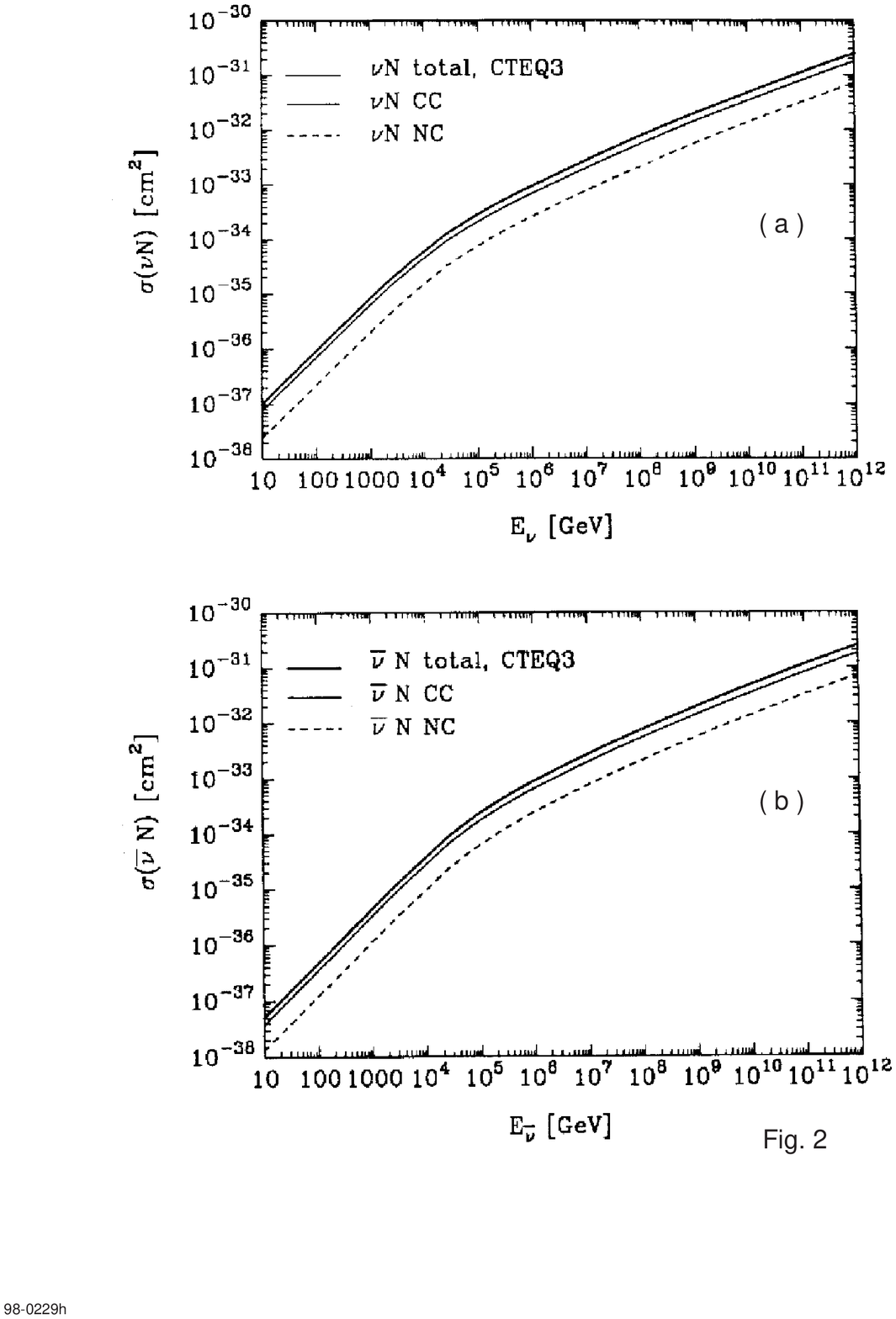,height=7in}
\item[] Fig. 2. Cross sections for (a) $\nu N$ interactions at high energies;
dotted line, $\sigma (\nu N \rightarrow \nu + {\rm anything})$; thin line,
$\sigma (\nu N \rightarrow \mu^- + {\rm anything})$; thick line, total
charged-current plus neutral-current) cross section. (b) for $\bar{\nu} N$
interactions. From reference 4, p. 90.
\newpage

\psfig{figure=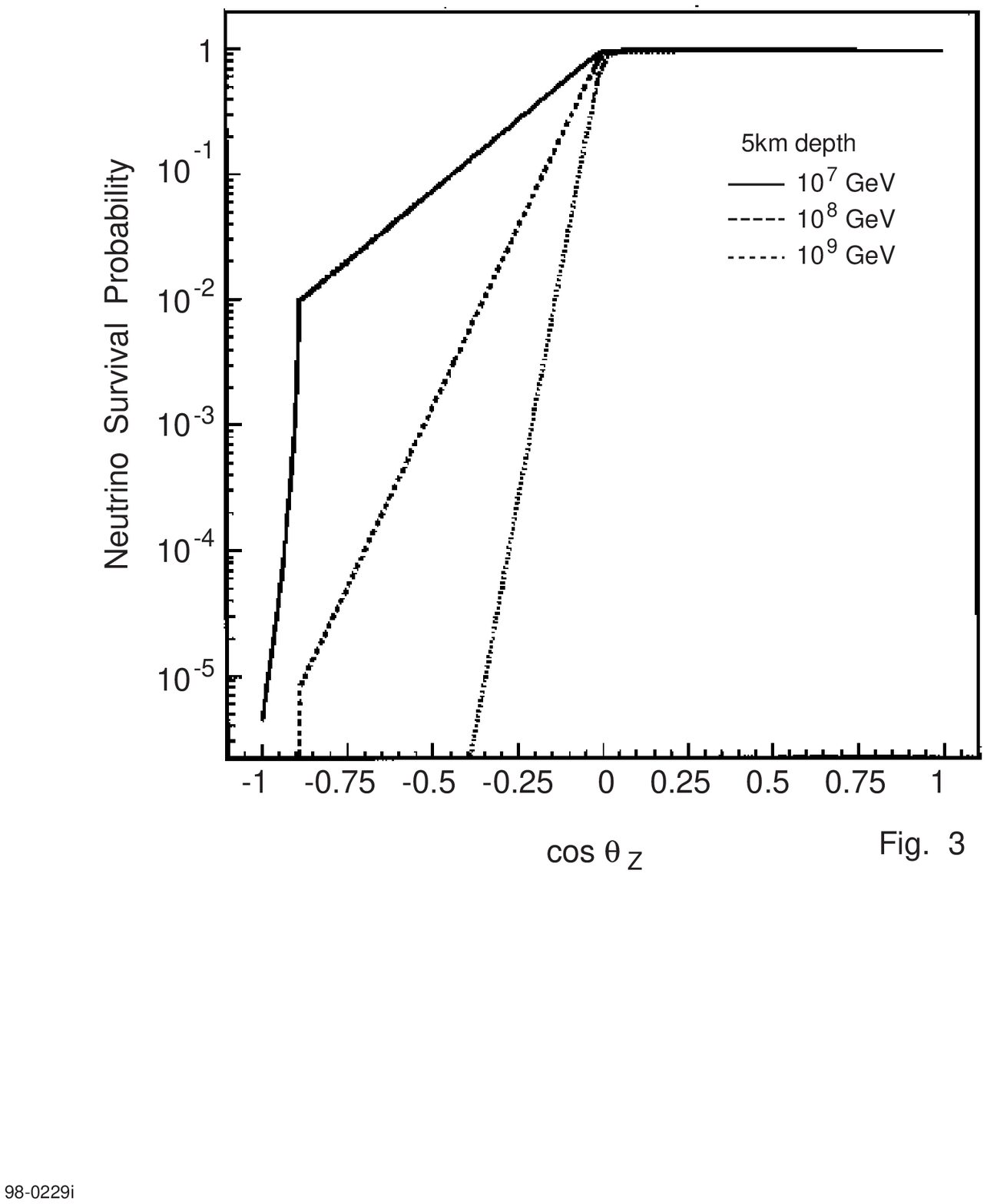,height=7in}
\item[] Fig. 3. Survival probability of UHE neutrinos from space as a
function of $\cos \theta_Z$, the zenith angle at a detector a few kilometers
below the Earth's surface; $\cos \theta_Z = +1$ corresponds to the zenith
direction.
\newpage

\hbox{
\psfig{figure=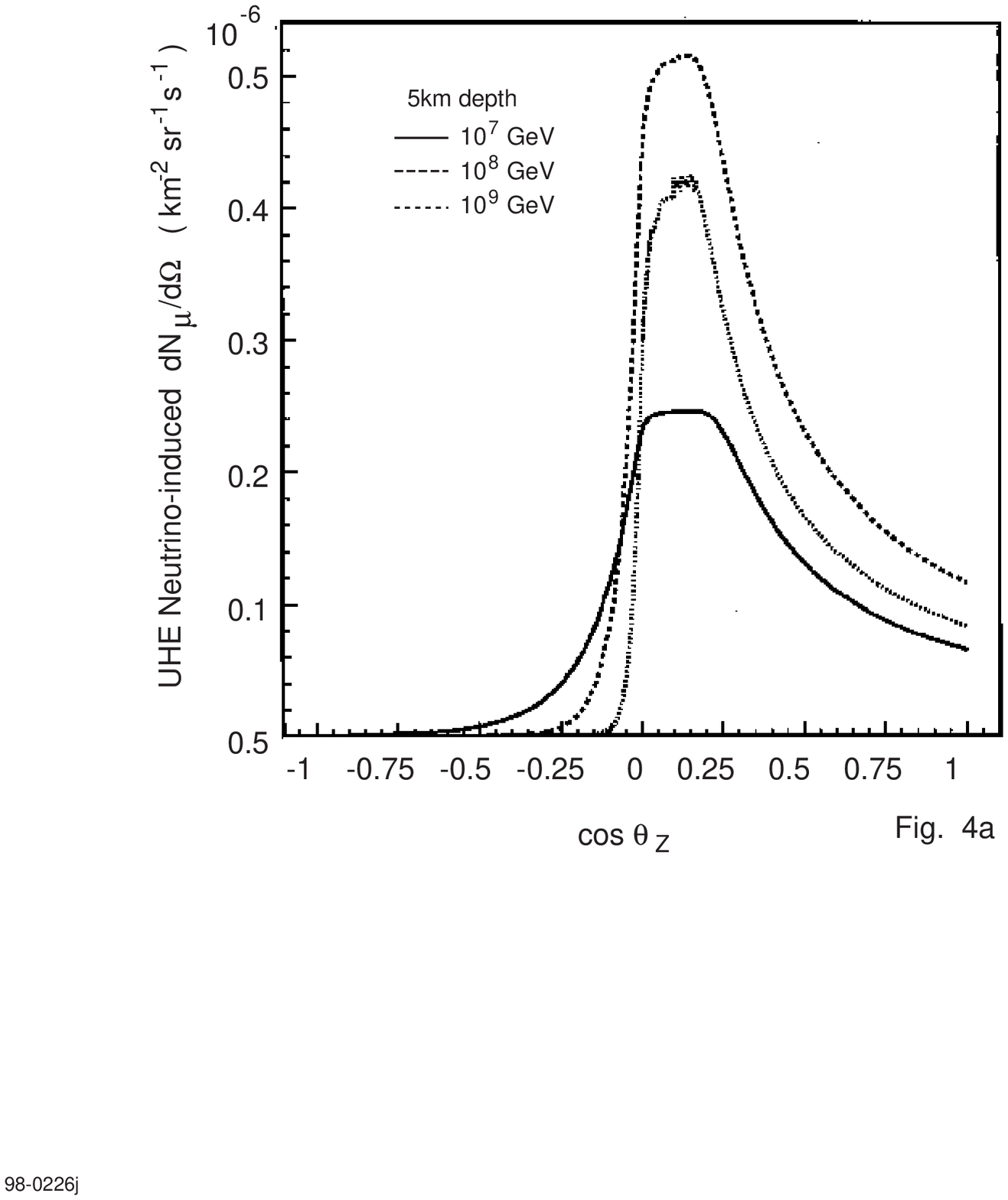,width=3.5in}
\psfig{figure=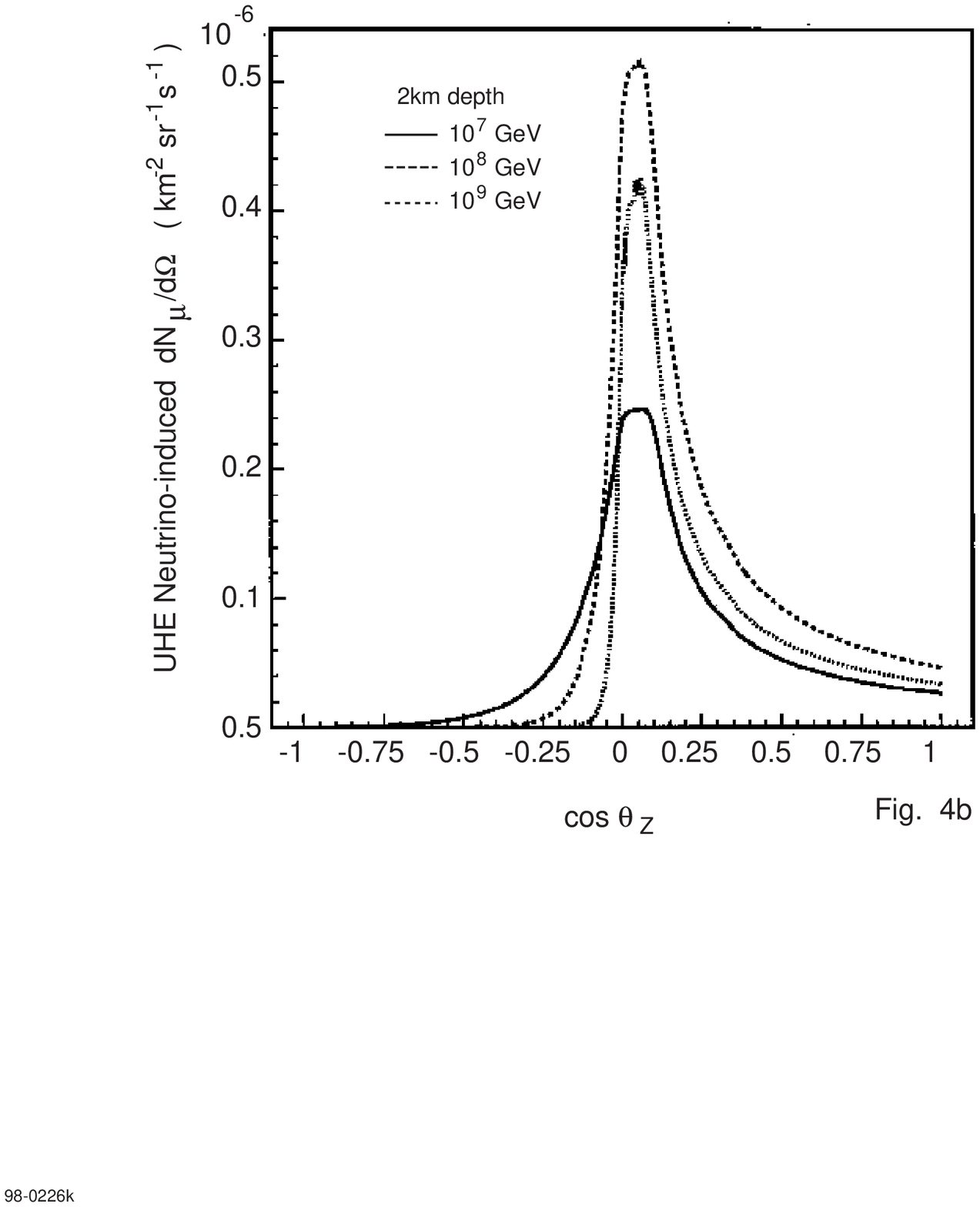,width=3.5in}
}
\item[] Fig. 4. Relative angular distributions of UHE neutrino-induced   muons
for three neutrino energies in an arbitrary detector located as in the Fig. 3
caption (a) detector 5 km deep, (b) detector 2 km deep.
\newpage

\psfig{figure=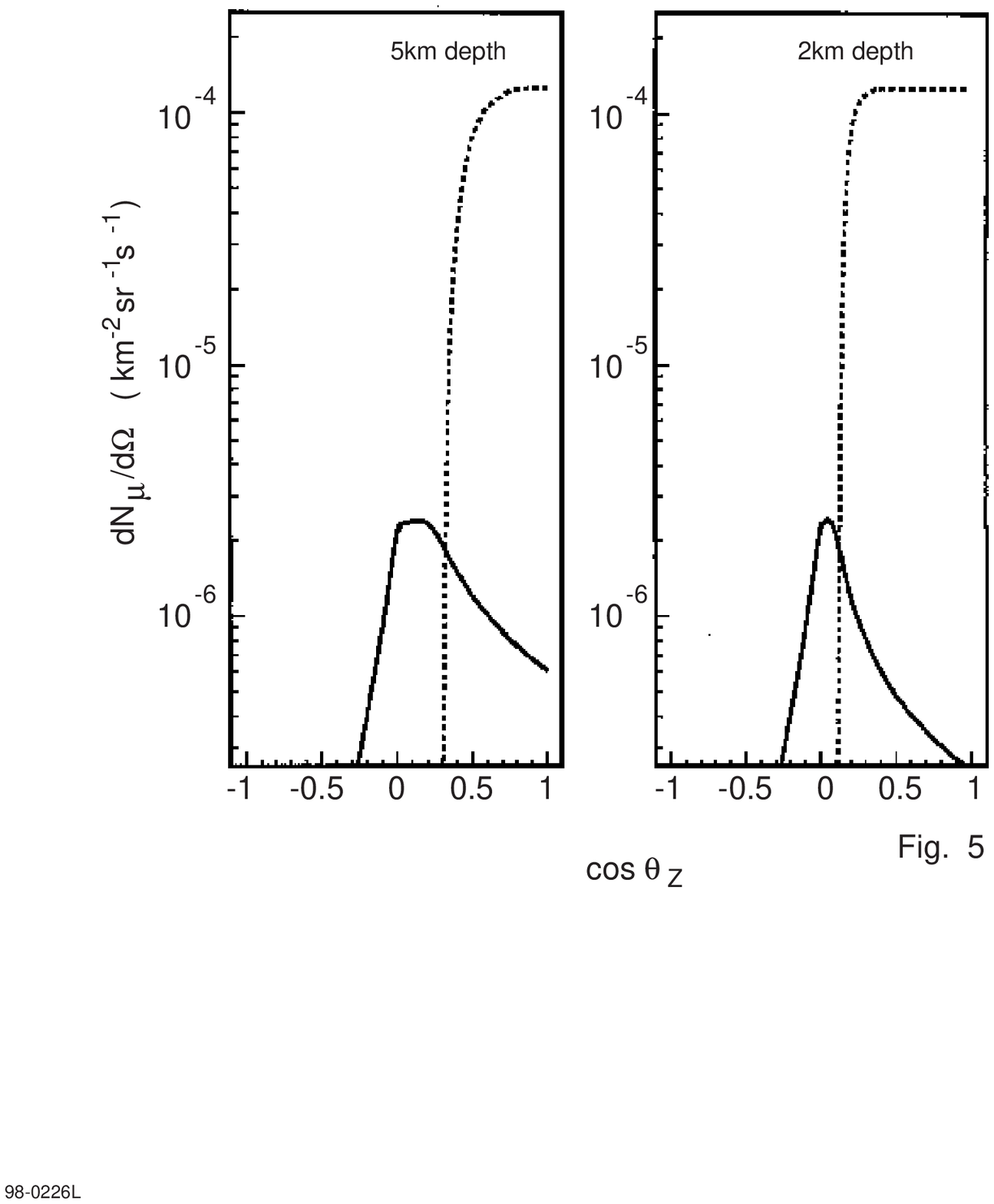,width=\textwidth}
\item[] Fig. 5. The angular distributions of UHE neutrino-induced muons
integrated over the energy interval $10^7$ to $10^{10}$ GeV with the cosmic
ray muon distributions superimposed (a) detector 5 km deep, (b) detector  2 km
deep.
\newpage

\psfig{figure=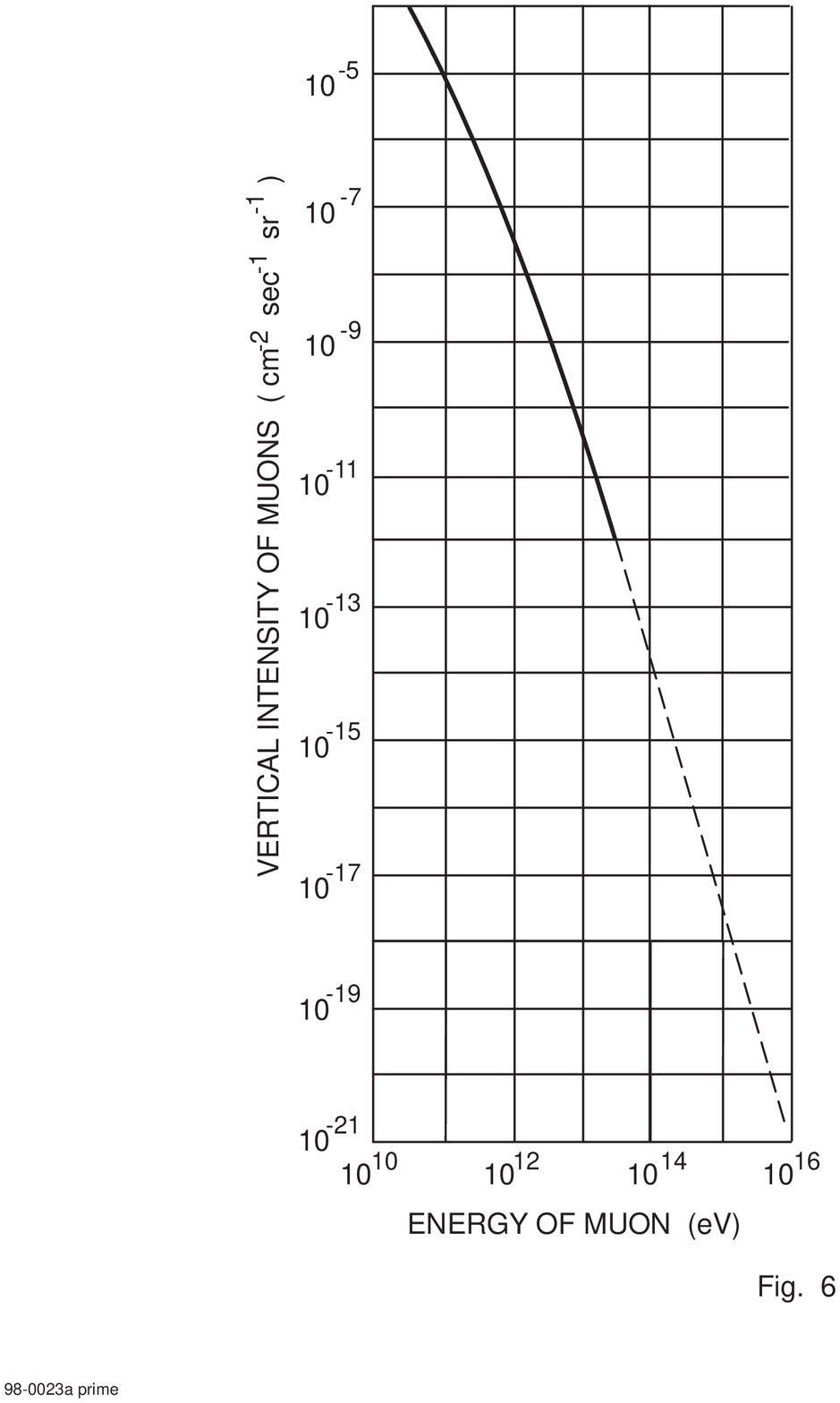,height=7in}
\item[] Fig. 6. Integral energy spectrum of cosmic ray
muons from reference 8, p. 401. The dashed line is the estimated extrapolation
of the original empirically based plot.
\newpage

\psfig{figure=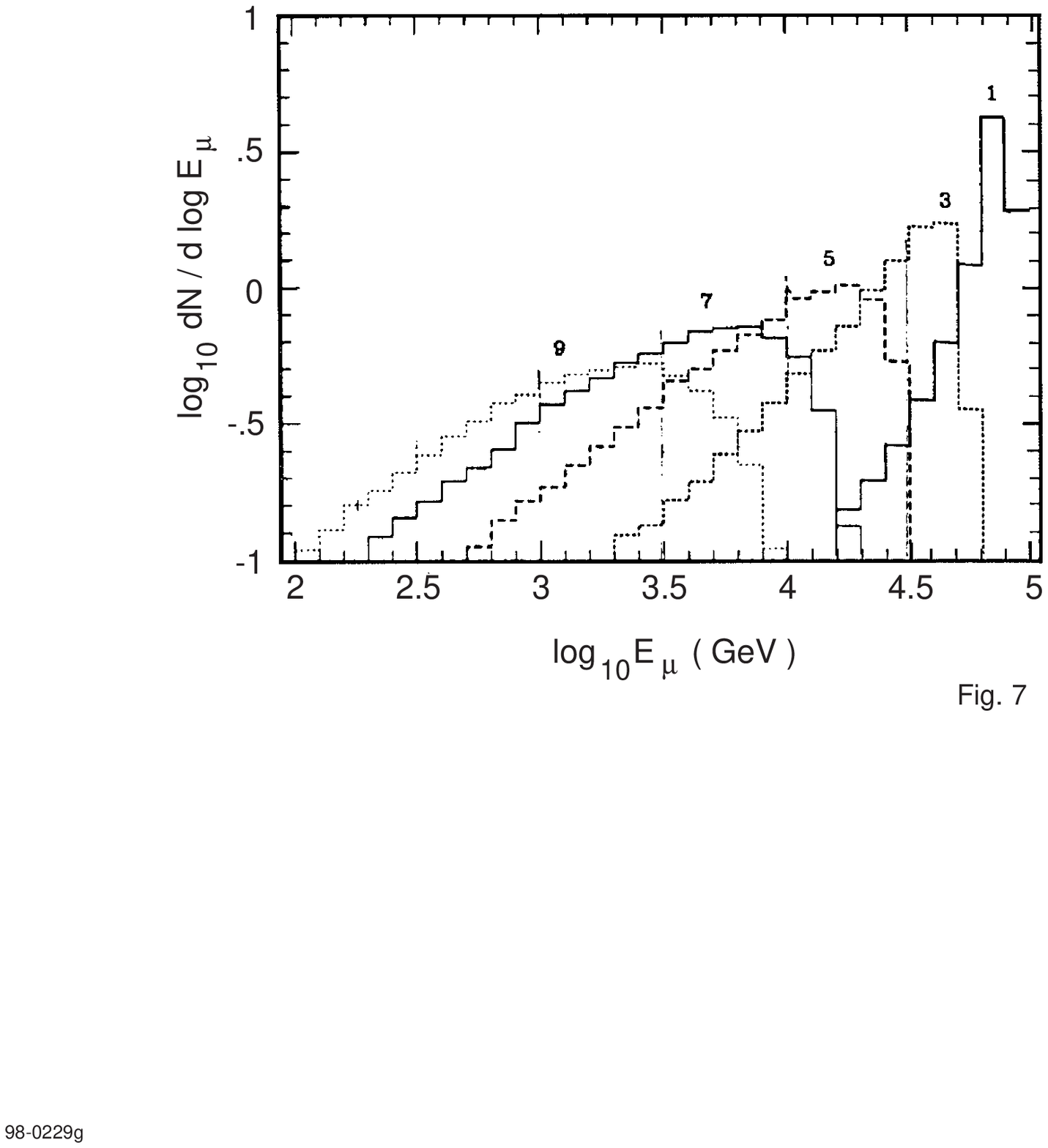,height=7in}
\item[] Fig. 7. Energy distribution of muons of energy $E_0 = 10^5$ GeV after
propagation in standard rock depths from 1 to 9 km w.e. The numbers by the
histograms show the corresponding depths. Reproduced from reference [8].
\newpage

\psfig{figure=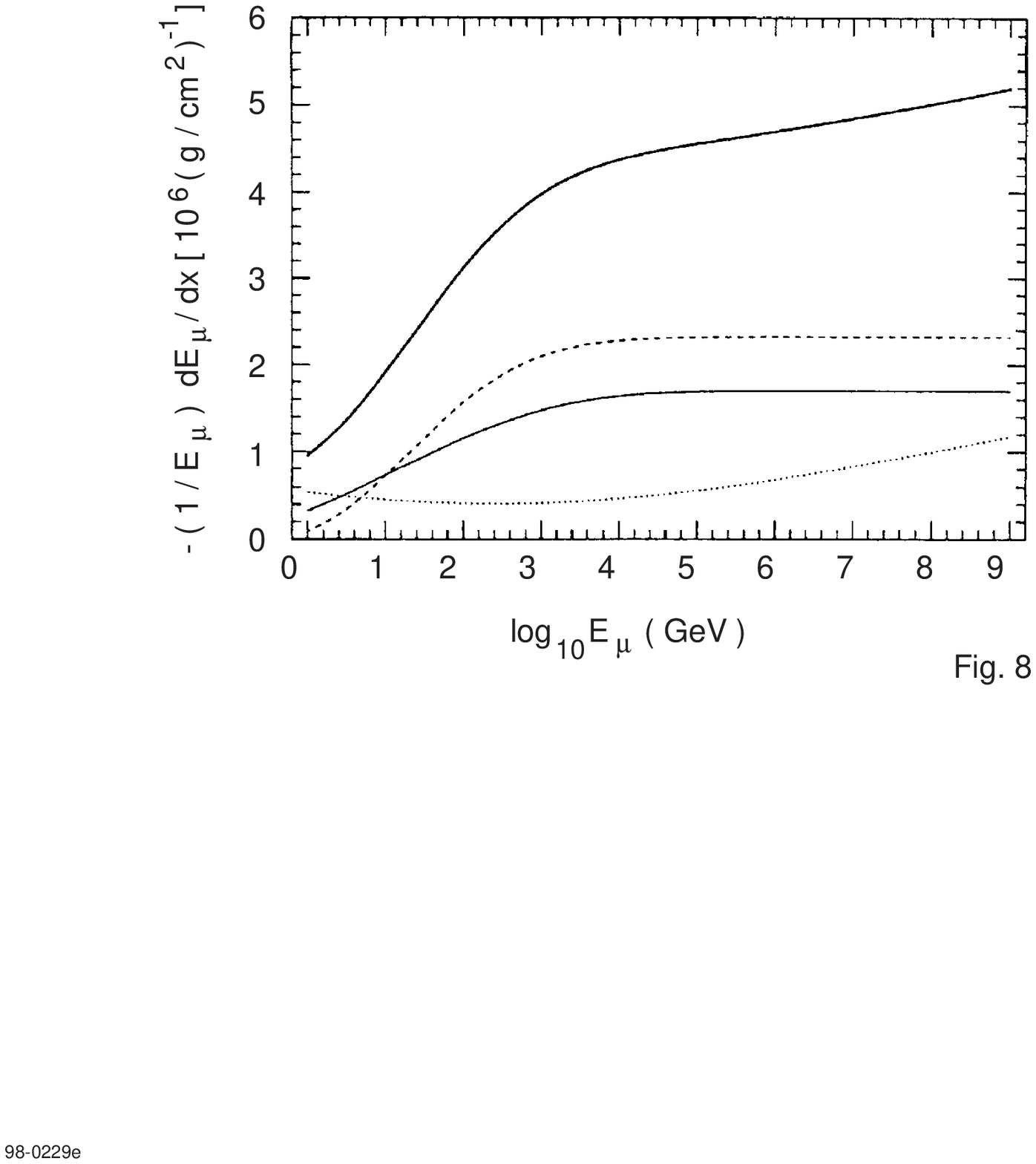,height=7in}
\item[] Fig. 8. Plot of the quantities $\beta_{rad} = {<dE/dx>}_{rad}/E$ for
the three radiative processes in standard rock as a function of muon energy E.
The solid line is for bremsstrahlung, dashed line for production of $e^+e^-$
pairs, and the dotted line for photoproduction. The thick solid line shows the
sum of the three processes. Reproduced from reference [8].

\end{enumerate}

\end{document}